# Quantum dots in suspended single-wall carbon nanotubes

Jesper Nygård[a)]
*Ørsted Laboratory, Niels Bohr Institute, Universitetsparken 5, DK-2100 Copenhagen Ø, Denmark*

David H. Cobden
*Department of Physics, University of Warwick, Coventry CV4 7AL, UK*

We present a simple technique which uses a self-aligned oxide etch to suspend individual single-wall carbon nanotubes between metallic electrodes. This enables one to compare the properties of a particular nanotube before and after suspension, as well as to study transport in suspended tubes. As an example of the utility of the technique, we study quantum dots in suspended tubes, finding that their capacitances are reduced owing to the removal of the dielectric substrate.

In most electrical transport experiments on single-wall nanotubes, the tubes are stuck to an insulating substrate (usually $SiO_2$) by van der Waals forces.[1] In spite of this, the electrical properties appear surprisingly similar to those expected for the ideal, unperturbed molecules.[2] Nevertheless the mechanical substrate-molecule interaction is strong enough to freeze in curves[3] and to squash tubes where they cross sharp features such as contact edges[4] or other tubes.[1] Other factors, such as the effects of dielectric polarization, and the contribution of trapped charges in the substrate to the observed disorder, hysteresis, and noise,[5] need yet to be understood.

It is therefore of interest both to directly investigate the influence of the substrate and to be able to measure the properties of tubes in the absence of a substrate. Freestanding tubes are also more suitable for exploring the interplay of electronic behavior with mechanical[6,7] and chemical properties, with possible applications in sensors,[8] transducers, and memory elements.[9] To this end, tubes have earlier been deposited[6,9] or grown[7] across prefabricated trenches. Here we demonstrate a very simple technique in which the tubes are first built into an electronic device and are then suspended by a wet oxide etch. The advantages of this procedure include simplicity, good anchoring of the suspended tube, and most importantly the ability to study transport properties of a particular tube both before and after substrate removal.

We begin by making standard nanotube devices. The tubes are grown either by laser vaporization[10] (these must be cut by sonication in dichloroethane and then deposited), or by chemical vapor deposition (CVD) directly on the substrate.[11,12] Electron-beam evaporated metallic electrodes consisting of Cr (~5 nm) followed by Au (~25 nm) are then patterned on top by electron-beam lithography using PMMA resist and lift-off. Fig. 1(a) shows a schematic cross-section of the resulting structure,[13] including the 300 nm oxide layer and the heavily doped silicon substrate which serves as a gate. At this point the device can be bonded into a package and characterised electrically.

We then immerse the silicon chip in a buffered $SiO_2$ wet etch (6.5% HF, Merck AF 87.5-12.5, 80 nm/min). The etch is terminated by transferring to water and then isopropanol, followed by drying in a nitrogen stream. The etch does not attack the gold or the carbon, leaving the electrodes raised and undercut, separated by a trench in the oxide. The Cr adhesion layer is essential to prevent the gold from washing away. Under certain circumstances the tubes remain suspended between the electrodes (Fig. 1(b)), while in others they end up draped down to the bottom of the trench.

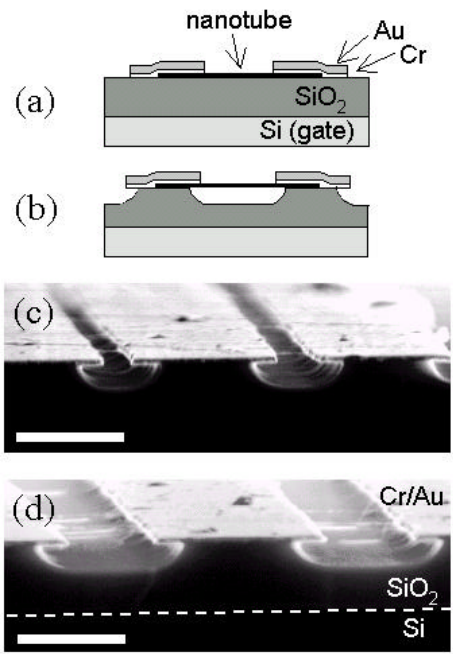

Figure 1: Above: processing of suspended nanotube devices. (a) Cross sectional schematic of a nanotube lying on a $SiO_2$/Si substrate and contacted by Cr/Au electrodes at each end. (b) After etching it is suspended between the raised electrodes. Below: scanning electron microscope images (JEOL JSM-6320F, 10 keV, x40,000) of test structures (150 nm etch) with electrode separations of (c) 150 nm and (d) 400 nm, cleaved for a cross-sectional view (scale bars 500 nm).

A number of factors influence the rate of successful suspension. These include the trench depth and width; the firmness of pinning to the electrodes; the straightness and thickness of the tubes or bundles; and the drying procedure. Tests with different electrode separation $d$ showed that the rate was much higher for $d = 150$ nm (Fig. 1(c)) than for $d = 400$ nm (Fig. 1(d)). Also, draping occurs for tubes which do not extend some distance under both electrodes, as well as in all cases for tubes deposited *on top* of pre-fabricated electrodes. This implies that to prevent the tubes slipping lengthwise to maximise their contact with the trench bottom, it is necessary to evaporate metal over several hundred nm of the ends of the tubes to pin them sufficiently tightly to the substrate. The suspension rate appeared higher for thicker bundles than for single tubes, presumably due to their higher rigidity. Nevertheless, suspended individual CVD-grown tubes were not difficult to obtain. Finally, although our simple drying procedure is adequate, it is possible that precautions for further reducing surface tension stress on drying[14] would also increase the success rate.

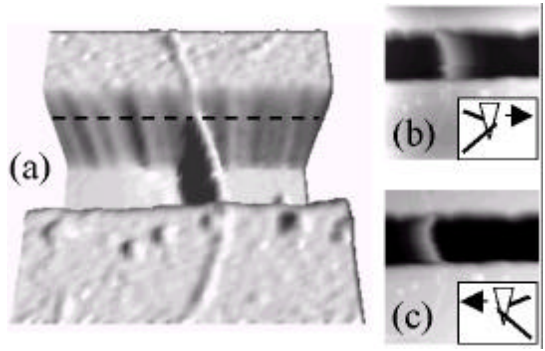

Figure 2: (a) AFM image of a 5 nm thick nanotube bundle suspended over a 50-nm deep trench of width $d = 300$ nm, taken in tapping mode at a scan speed of 1 μm/s. We used a Nanoscope III with a Si cantilever, stiffness 42 N/m (vertical), >1000 N/m (lateral). A dashed line indicates the metal-$SiO_2$ boundary. (b) Left-to-right and (c) right-to-left scans at 20 μm/s. Insets indicate how the bundle is pulled in the scan direction.

The suspended tubes were imaged using an atomic force microscope (AFM) in tapping mode. Fig. 2(a), taken at a slow scan speed (1 μm/s), shows a 5-nm thick bundle of laser-ablation grown tubes bridging a trench. Its profile can be seen beneath the metal at either end, and the suspended part appears straight, implying that it is taut. However, at a faster scan speed of 20 μm/s it appears bowed in the scan direction (Figs. 2(b) and (c)). This is because the finite response time of the AFM causes the suspended bundle to be stretched some distance in the scan direction before the tip retracts. The observed distortion corresponds to a strain in the bundle of no more than 5 %. The actual strain is probably less than this, taking into account possible slip underneath the metal and distortion of the cantilever. Similar elastic deformation of suspended tubes by contact-mode AFM was used earlier by Walters et al[14] to measure their elastic properties, and by Tombler et al[7] to observe electromechanical coupling.

The devices survive experiments in liquid helium. Fig. 3 presents the characteristics of an individual suspended CVD-grown nanotube at 4.2 K. The linear conductance $G$ vs gate voltage $V_g$ exhibits regular Coulomb blockade (CB) oscillations (Fig. 3(a)), implying that a quantum dot is formed in the suspended tube. A grayscale plot (Fig 3(b)) of the differential conductance $dI/dV$ against $V_g$ and source-drain bias $V$ reveals Coulomb diamonds[15] with just-resolvable excited states (parallel dark lines). From the height of the diamonds we deduce a charging energy $U \approx 5$ meV and from the excitation spectra a level spacing $\Delta \sim 1$ meV. These are typical for a nanotube quantum dot of length $L$ of the order 500 nm,[13] and the ratio $U/\Delta$ is consistent with the value $\sim 6$, typically found for tubes on substrates.[16]

Our procedure also gives us the ability to directly compare the electronic properties of a particular tube with and without the substrate. The inset to Fig. 4(a) is an image of a 4 nm bundle suspended in three sections. Fig. 4(a) shows $G$ vs $V_g$ at a series of $T$ for the lower section *before* etching. All three sections showed similar characteristics, which are typical for a metallic tube with tunneling contacts.[16] Fig. 4(b) shows the characteristics of the same section *after* etching the substrate. The CB oscillations remain, with an approximately unchanged charging energy of roughly $U \approx 8$ meV, implying that the same quantum dot is still present in the suspended bundle. Furthermore, the average height of the oscillations is also unchanged, implying that contacts between the dot and the leads were little affected by the etch.

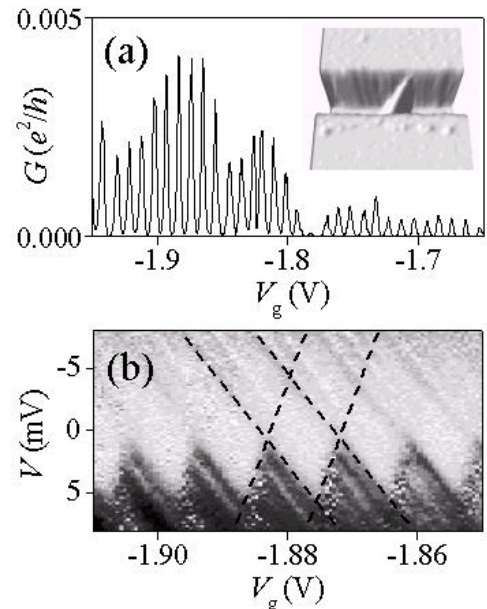

Figure 3: (a) Coulomb blockade oscillations in a suspended CVD-grown nanotube at 4.2 K. Inset: AFM image of such a tube, with diameter 1.5 nm and electrode separation $d = 250$ nm. (b) Grayscale plot of $dI/dV$ (darker = more positive). The dashed lines demark a Coulomb diamond.

One major change is however apparent on etching: the period of the CB oscillations, $\Delta V_g$, increases dramatically, from about 70 mV to 160 mV. Since $\Delta V_g \approx e/C_g$, this means that the capacitance $C_g$ between the dot and the gate decreases, just as one would expect as a consequence of removing part of the dielectric ($\varepsilon_r = 3.9$ for $SiO_2$) between the bundle and the gate. The capacitances $C_s$ and $C_d$ between the dot and the source and drain electrodes are much less affected. Since $U = e^2/(C_g + C_s + C_d)$, we have $e\Delta V_g/U = 1 + (C_s + C_d)/C_g$. In all our devices $e\Delta V_g/U \gg 1$, so $C_s + C_d \gg C_g$ and $U \approx e^2/(C_s + C_d)$. As we see no noticeable decrease in $U$, we conclude that $C_s$ and $C_d$ are not greatly changed by the etching. This can only be explained if $C_s$ and $C_d$ are determined mainly by the parts of the tube underneath the electrodes.

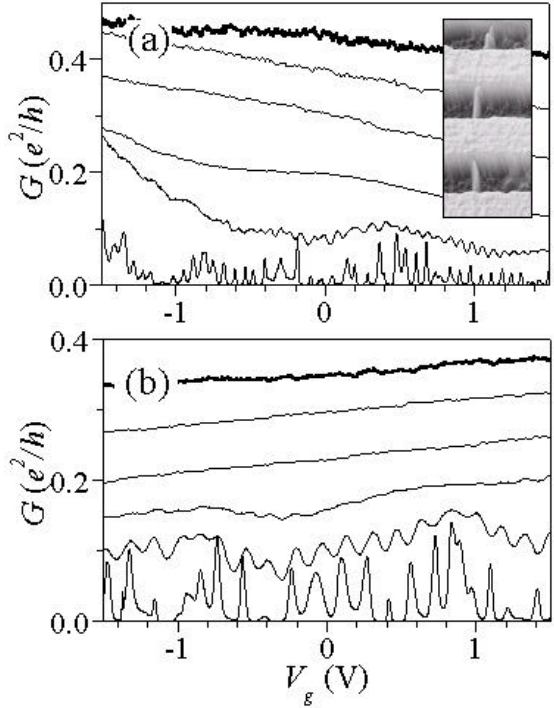

Figure 4: Inset: AFM perspective image of a 4 nm bundle bridging a series of four electrodes separated by trenches of width 300 nm and depth 200 nm. (a) Conductance vs gate voltage for the lower section *before* suspending this bundle. The sample was immersed in helium at temperatures $T = 300$ (thick trace), 175, 110, 50, 23, and 4.2 K. (b) $G$ vs $V_g$ *after* suspending the bundle. Same $T$ sequence as in (a).

Even so, for larger electrode spacings, $e\Delta V_g/U$ has been found to approach unity,[17] implying that $C_g$ dominates $C_s$ and $C_d$. Thus for longer suspended tubes the change of $C_g$ should result in a large increase in $U$. Since $\Delta = hv_F/4L$ depends only on the length $L$ of the dot, one then expects $U/\Delta$ to increase on etching. The quantity $U/\Delta$ characterises the strength of long-range electron-electron interactions in the tube, which are thought to be the cause of the commonly observed steady decrease of $G$ on cooling, as seen in Fig. 4. Although the behavior is consistent with the predicted formation of a Luttinger liquid within the tube,[16] it is difficult to conclusively separate the various effects which can suppress tunneling into a small conductor at low energies. The unique ability to tune the interaction strength may prove decisive in this regard.

Trapped charges in the substrate are obvious candidate sources of the noise and hysteresis seen in all nanotube devices around room temperature, which will hinder their applications in electronics. Surprisingly we find no consistent improvement in etched devices, hinting that these problems may be related more to adsorbates or contact defects than to the substrate.

The simplicity and robustness of this technique lend it many more potential uses. Suspended tubes are suitable for investigating their vibrational, thermal, electromechanical and electrochemical properties, and their coupling to a wide range of environments. We have also performed tests to show that our suspended nanotubes can act as templates for evaporating metal nanowires,[18] with the etched undercut preventing shorting by the metal deposited on the bottom of the trench.

We thank R. Fléron for the key idea, P. R. Poulsen for CVD processing, and M. M. Andreasen, T. Hassenkam, and N. Wilson for experimental help. The laser-ablation nanotubes were supplied by A. G. Rinzler. The work was carried out in the NBI Nanolab and supported by the Danish Technical Research Council.